\documentclass{article} %
\usepackage{iclr2020_conference, times}

\usepackage{amsmath,amsfonts,bm}

\def\eqref#1{equation~\ref{#1}}

\def\1{\bm{1}}

\def\vh{{\bm{h}}}

\def\vx{{\bm{x}}}

\DeclareMathAlphabet{\mathsfit}{\encodingdefault}{\sfdefault}{m}{sl}
\SetMathAlphabet{\mathsfit}{bold}{\encodingdefault}{\sfdefault}{bx}{n}

\usepackage{hyperref}
\usepackage{url}

\title{SCELMo: Source Code Embeddings \\ from Language Models}

\author{Rafael~-~Michael ~Karampatsis \\ 
Department of Computer Science\\
The University of Edinburgh\\
Edinburgh, EH8 9AB, UK \\
\texttt{r.m.karampatsis@sms.ed.ac.uk} \\
\And
Charles ~Sutton \\
Google Research, The University of Edinburgh \\
Mountain View, CA, United States\\
\texttt{charlessutton@google.com} \\
}

\usepackage{listings}
\usepackage{color}
\definecolor{lightgray}{rgb}{.9,.9,.9}
\definecolor{darkgray}{rgb}{.4,.4,.4}
\definecolor{purple}{rgb}{0.65, 0.12, 0.82}

\lstdefinelanguage{JavaScript}{
  keywords={typeof, new, true, false, catch, function, return, null, catch, switch, var, if, in, while, do, else, case, break},
  keywordstyle=\color{blue}\bfseries,
  ndkeywords={class, export, boolean, throw, implements, import, this},
  ndkeywordstyle=\color{darkgray}\bfseries,
  identifierstyle=\color{black},
  sensitive=false,
  comment=[l]{//},
  morecomment=[s]{/*}{*/},
  commentstyle=\color{purple}\ttfamily,
  stringstyle=\color{red}\ttfamily,
  morestring=[b]',
  morestring=[b]"
}

\usepackage{booktabs}
\usepackage[margin=1in]{geometry}

\usepackage{xspace}
\newcommand{\wVec}{Word2Vec\xspace}
\newcommand{\DB}{DeepBugs\xspace}
\newcommand{\ft}{FastText\xspace}
\newcommand{\js}{JavaScript\xspace}

\newcommand{\cs}[1]{}
\iclrfinalcopy %
\begin{document}
\maketitle

\begin{abstract}
Continuous embeddings of tokens in computer programs have been used to support a variety
of software development tools, including readability, code search, and program repair.
Contextual embeddings are common in natural language processing but have not been previously applied
in software engineering.
We introduce a new set of {deep contextualized} word representations for computer programs
based on language models.
We train a set of embeddings using the ELMo (embeddings from language models) framework of Peters et al (2018).
We investigate whether these embeddings are effective when fine-tuned for the downstream task of bug detection.
We show that even a low-dimensional embedding trained on a relatively small corpus of programs
can improve a state-of-the-art machine learning system for bug detection.
\end{abstract}

\section{Introduction} \label{intro}
Learning rich representations for source code is an open problem that has the potential
to enable software engineering and development tools.
Some work on machine learning for source code has used hand engineered features \citep[e.g.]{Long2016},
but designing and implementing such features can be tedious and error-prone.
For this reason, other work considers the task of learning
 a representation of source code from data \citep{Allamanis2018}. 
Many models of source code are based on learned representations  called embeddings, which transform words 
into a continuous vector space \citep{Mikolov2013}.
Currently in software engineering (SE) 
researchers have used static embeddings \citep{Harer2018, White2019, Pradel2018},
which map a word to the same vector regardless of its context.
However, recent work in natural language processing (NLP) has found that contextual embeddings 
can lead to better performance \citep{Peters2018, Devlin2018, Yang2019, Liu2019}.
Contextualized embeddings assign a different vector to a word based on the context it is used.
For NLP this has the advantage that it can model phenomena like polysemy.
A natural question to ask is if these methods would also be beneficial for learning better SE representations.

In this paper, we introduce a new set of contextual embeddings for source code. Contextual embeddings have several potential modelling advantages that are specifically suited to modelling source code:
\begin{itemize}
\item Surrounding names contain important information about an identifier. For example, for a variable name, 
surrounding tokens might include functions that take that variable as an argument or assignments to the variable. 
These tokens provide indirect information about possible values the variable could take,
and so should affect its representation.
Even keywords can have very different meanings based on their context.
For instance, a private function is not the same as a private variable or a private class (in the case of Java / C++).
\item Contextual embeddings assign a different representation to a variable each time it is used in the program.
By doing this, they can potentially capture how a variable's value evolves through the program execution.
\item Contextual embeddings enable the use of transfer learning.
Pre-training a large neural language model and querying it for contextualized representations while simultaneously fine-tuning for the specific task is a very effective technique for supervised tasks for which there is a small amount 
of supervised data available.
As a result only a small model needs to be fine-tuned atop the pre-trained model, without the need for task-specific architectures nor the need of training a large model for each task separately.
\end{itemize}

In this paper, we highlight the potential of contextual code embeddings for program repair.
Automatically finding bugs in code is an important open problem in SE.
Even simple bugs can be hard to spot and repair.
A promising approach to this end is name-based bug detection, introduced by \DB\ \citep{Pradel2018}.
The current state-of-the-art in name-based bug detection relies on static representations from \wVec\ \citep{Mikolov2013} to learn a classifier that distinguishes correct from incorrect code for a specific bug pattern.
We introduce a new set of contextualized embeddings for code and explore its usefulness on the task of name-based bug detection.
Our method significantly outperforms \DB\ as well as other static representations methods on both the \DB dataset as well as a new previously unused test set of \js\ projects.

\section{Related Work}

Unsupervised static word embeddings have been extensively used to improve the accuracy of supervised tasks in NLP \citep{Turian2010}.
Notable examples of such methods are Word2Vec \citep{Mikolov2013} and GloVe \citep{Pennington2014}.
However, the above models learn only a single context-independent word representation.
To overcome this problem some models \citep{Wieting2016, Bojanowski2017} enhance the representations with subword information, which can also somewhat deal with out-of-vocabulary words.
Another approach is to learn a different representation for every word sense \citep{Neelakantan2014} but this requires 
knowing the set of word senses in advance.
More recent methods overcome the above issues by learning contextualized embeddings.
\cite{Melamud2016} encode the context surrounding a pivot word using a bidirectional LSTM.
\cite{Peters2018} use a deep bidirectional LSTM, learning word embeddings as functions of its internal states,
calling the method Embeddings using Language Models (ELMo). We discuss ELMo in detail in Section~\ref{sec:elmo}.
\cite{Devlin2018} introduced bidirectional encoder representations from transformers (BERT).
This method learns pre-trained contextual embeddings by jointly conditioning on left and right context via an attention mechanism.

Program repair is an important task in software engineering and programming languages.
For  a detailed review see \citet{Monperrus2018, Gazzola2019}.
Many recent program repair methods are based on machine learning.
\cite{Yin2018} learn to represent code edits using a gated graph neural network (GGNN) \citep{Li2016gated}.
\cite{allamanis2018learning} learn to identify a particular class of bugs called variable misuse bugs,
using a GGNN.
\cite{Chen2019Sequencer} introduce SequenceR which learns to transform buggy lines into fixed ones via machine translation. 
Our work is orthogonal to these approaches and can be used as input in other models.

Finally, our work is also related to code representation methods many of which have also been used in program repair.
\cite{Harer2018} learn \wVec\ embeddings for C/C++ tokens to predict software vulnerabilities.
\cite{White2019} learn \wVec\ embeddings for Java tokens and utilize them in program repair.
\cite{Alon2019} learn code embeddings using abstract syntax tree paths.
A more detailed overview can be found in \citep{Allamanis2018, Chen2019}.

\section{Embeddings from Language Models (ELMo)} \label{sec:elmo}

ELMo \citep{Peters2018} computes word embeddings from the hidden states of a language model.
Consequently, the embeddings of each token depend on its context of the input
sequence,  even out-of-vocabulary (OOV) tokens have effective input representations.
In this section, we briefly describe the ELMo embeddings.

The first step is that a neural language model is trained to maximize the likelihood of a training corpus.
The architecture used by ELMo a bidirectional LSTM with $L$ layers and character convolutions in 
the input layer.
Let the input be
a sequence of tokens $(t_1, ... t_N).$
For each token $t_k$, denote by $\vx_k^{LM}$  the input representation from the character convolution.
Consequently, this representation passes through $L$ layers of forward and backward LSTMs.
Then each layer $j \in \{1, ..., L\}$ of the forward LSTM computes a hidden state $\overrightarrow{\vh_{k,j}^{LM}}$, 
and  likewise the hidden states of the backward LSTM are denoted by $\overleftarrow{\vh_{k,j}^{LM}}$.
The parameters for the token representation and for the output softmax layer are tied for both directions, while different parameters are learned for each direction of the LSTMs.

After the language model has been trained, we can use it within another downstream 
task by combining the hidden states of the language model from each LSTM layer. This process is called ELMo.
For each token $t_k$ of a sentence in the test set,
the language model computes $2L + 1$ hidden states, one in each direction for each layer,
and then the input layer.
To make the following more compact, we can write these as
$h_{k,0}^{LM} = x_k^{LM}$ for the input layer, and then
$h_{k,j}^{LM} = \lbrack \overrightarrow{h_{k,j}^{LM}}, \overleftarrow{h_{k,j}^{LM}}\rbrack$
for all of the other layers. The set of these vectors is
\begin{equation} \label{eq:concat}
R_k = \lbrace h_{k,j}^{LM} | j = 0, ..., L \rbrace.
\end{equation} 
To create the final representation that is fed to downstream tasks, ELMo collapses the set of representations into a single vector $E_k$ for token $t_k$.
A simplistic approach is to only select the top layer, so that $E_k = h_{k,L}^{LM}$.
A more general one, which we use in this work, 
is to combine the layers via fine-tuned task specific weights $\mathbf{s} = (s_1 \ldots s_L)$ for every layer.
Then we can compute the embedding for token $k$ as
\begin{equation} \label{elmo}
E_k = \gamma \sum_{j=0}^{L}s_j h_{k,j}^{LM},
\end{equation}
where $\gamma$ is an additional scalar parameter that scales the entire vector.
In our experiments we did not performed fine-tuning and thus used equal weights $s_j = 1/(L+1)$ for each layer and $\gamma = 1$.
However, our implementation also supports all the aforementioned ways of collapsing the set of representations.

A potential drawback of the method is that it still utilizes a softmax output layer with a fixed vocabulary that does not scale effectively and it still predicts UNK for OOV tokens which may have a negative effect on the representations.

\section{Source Code ELMo} \label{data}

We describe Source Code ELMo (SCELMo), which trains ELMo on corpora of source code.
However, we note that normally ELMo models in other domains are able to effectively utilize much larger representations. %
The code was tokenized using the esprima \js\ tokenizer\footnote{\url{https://esprima.org/}}.
\cs{We tokenize according to the Javascript tokenizer?}
For training the ELMo model we used a corpus of 150,000 \js\ Files (Raychev et al. 2016) consisting of various open-source projects.
This corpus has previously been used on several tasks \citep{Raychev2016, Pradel2018, Bavishi2018}.
We applied the patch released by \cite{Allamanis2018} to filter out code duplication as this phenomenon was shown on this and other corpora to result in inflation of performance metrics.
This resulted in 64750 training files and 33229 validation files.
Since the validation set contains files from the same projects as the train the contained instances might be too similar and unrealistic overestimating.
To address this we also created a test set of 500 random \js\ projects sampled from the top 20,000 open-source \js\ projects as of May 2019.
The test corpus has not been previously utilized in previous work and is a better reflection of the performance of the learned bug detectors.
Lastly, it is important to know what the performance of the method will be if we do not have access to training data from the projects on which we would like to find bugs.
This is common in practice for many real case scenarios.
For training the ELMo model, we use an embedding size of 100 features for each of the forward and backward LSTMs so that each layer sums up to 200 features.

\section{Contextual Embeddings for Program Repair} \label{method}

In this section, we describe how contextual embeddings can be incorporated within
a recent machine learning-based bug detection system, the \DB\ system of
\citet{Pradel2018}. In the first part of this section, we give background about the \DB\ system,
and then we describe how we incorporate SCELMo within \DB.
\DB treats the problem of finding a bug as a classification problem.
The system considers a set of specific bug types, which are small mistakes
that might be made in a program, such as swapping two arguments.
For each bug type, \DB\ trains a binary classifier that takes a program statement
as input and predicts whether the statement contains that type of bug.
At test time, this classifier can be run for every statement in the program
to attempt to detect bugs.

In order to train the model both examples of correct and incorrect (buggy) code are necessary.
\DB treats the existing code as correct and randomly mutates it to obtain buggy code.
To obtain training examples, we extract all expressions from the source code which are either 
the function calls with exactly two arguments and all binary expressions.
To create instances of buggy code we mutate each of the correct instances.
As such, arguments in function calls are swapped, the binary operator in binary expressions is replaced with another random one, and finally randomly either the left or the right operand is replaced by another random binary operand that appears in the same file. Then the classification task is a binary task to predict whether the instance is correct, i.e., it comes
from the original code, or whether it is buggy, i.e. it was one of the randomly mutated examples.
The validation and test sets are mutated in the same way as the training set.
The split between correct and buggy instances has 50/50 class distribution as for each original code instance exactly one mutated buggy counterpart is created.

The architecture for the classifier is a feedforward network with a single hidden layer of 200 dimensions with Relu activations and a sigmoid
output layer.
For both the input and hidden layers a dropout of 0.2.
The network was trained in all experiments for 10 epochs with a batch size of 50 and the RMSProp optimizer.
We note that for maintaining a consistent comparison with \DB\ we kept all the above parameters as well as the optimizer's parameters fixed to the values reported in \citet{Pradel2018}.
Tuning these parameters would probably result in at least a small performance increase for our method.

\begin{figure}[h]
\begin{center}

\begin{lstlisting}[caption=Swapped Arguments Bug]
// Argument order is inversed.
var delay = 1000;
setTimeout(delay, function() { // Function should be first.
    logMessage(msgValue);     
});
\end{lstlisting}

\begin{lstlisting}[caption=Incorrect Binary Operator]
// && instead of || was used.
var p = new Promise();
if (promises === null && promises.length === 0) {
	p.done(error, result);
}
\end{lstlisting}

\begin{lstlisting}[caption=Incorrect Binary Operand]
// Call to .length is missing.
if ( index < matrix ) {
	do_something();
}
\end{lstlisting}
\end{center}
\caption{Bug type examples.}
\label{fig:patternExamples}
\end{figure}

In our experiments, we consider three bug types
that address a set of common programming mistakes:
swapped arguments of function calls, using the wrong binary operator and using an incorrect binary operand in a binary expression.
The methodology can easily be applied to other bug types.
Figure~\ref{fig:patternExamples} illustrates an example of each of the three bug types.

\subsection{Input to the Classifier}

A key question is how a statement from the source code
is converted into a feature vector that can be used within the classifier.
\DB\ uses a set of heuristics that, given a statement and a bug type, 
return a sequence of identifiers from the statement that are most likely
to be relevant. 
For instance, for the call to setTimeout in Listing~1 the following sequence of identifiers would be extracted: \textit{[setTimeout, delay, function]}.
A detailed description of the heuristics is available in Appendix~\ref{sec:extraction}.
\cs{Rafael can you add an example here. Say what the tokens are for one of the listing.}
\cs{Add forward pointer to the appendix.}

These heuristics result in a sequence of program identifiers. These are converted to continuous vectors
using word embeddings, concatenated, and this is the input to the classifier. 
\DB\ uses \wVec embeddings trained on a corpus of code. 
In our experiments, we train classifiers using three different types of word embeddings.
First, we kept the 10,000 most 
frequent identifiers/literals and assigned to each of them a \emph{random embedding} of 200 features.
Second, to reproduce the results of \citet{Pradel2018},
we use the CBOW variant of \emph{Word2Vec} to learn representations consisting of 200 features for the 10,000 most frequent identifiers/literals.
Finally, we train a \emph{\ft} embeddings \citep{Bojanowski2017} on the training set to learn identifier embeddings that contain subword information.
The subwords used by \ft\ are all the character trigrams that appear in the training corpus.
Identifiers are therefore composed of multiple subwords.
To represent an identifier, we sum the embeddings of each of its subwords and summing them up.
This allows the identifier embeddings to contain information about the structure and morphology of identifiers.
This also allows the \ft embeddings, unlike the \wVec\ ones, to represent OOV words as a combination
of character trigrams. %

Note that DeepBugs can detect bugs only in statements that do not contain OOV (out-of-vocabulary) identifiers,
because its Word2Vec embeddings cannot extract features for OOV names.
Instead our implementation does not skip such instances.
Since the original work discarded any instances that contain OOV identifiers we neither know how the method performs on such instances nor how often those appear in the utilized dataset of DeepBugs.
Moreover, DeepBugs supported only a specific subset of AST nodes and skipped the rest.
For example if a call's argument is a complex expression consisting of other expressions then the call would be skipped.
However, we expanded the implementation to support all kinds of AST nodes and to not skip instances with nested expressions as discussed in Appendix~\ref{sec:extraction}.
We note that we still skip an instance if one of its main parts (e.g., a function call's argument) is a complex expression longer than 1,000 characters as such expressions might be overly long to reason about.

\subsection{Connecting SCELMo to the Bug Detector}

We investigated two variants of the bug detection model, which query SCELMo in
different ways to get features for the classifier.
The first utilizes the heuristic of Section~\ref{sec:extraction} to extract a small set of identifiers or literals that represent the code piece.
For example, for an incorrect binary operand instance we extract one identifier or literal for the left and right operands respectively, and we also extract its binary operator.
Then, those are concatenated to form a query to the network.
In the case of function calls we extract the identifier corresponding to the name of the called function, one identifier or literal for the first and second argument respectively and an identifiers for the expression on which the function is called. 
We also add the appropriate syntax tokens (a '.' if necessary, ',' between the two arguments, and left and right parentheses) to create a query that resembles a function call.
This baseline approach creates simplistic fixed size queries for the network but does not utilize its full potential since the queries do not necessarily resemble  actual code, nor correct code similar to the sequences 
in the training set for the embeddings.
We will refer to this baseline as No-Context ELMo.

Our proposed method, we compute SCELMo embeddings
to the language model all the tokens of the instances for which we need representations.
Valid instances are functions calls that contain exactly two arguments and binary expressions.
\cs{What does instances mean, the entire source file? One method?}
To create a fixed-size representation we extract only the features corresponding a fixed set of tokens.
Specifically, for functions calls we use the representations corresponding to the first token of the expression on which the function is called, the function name, the first token of the first argument and the first token of the second argument.
While, for binary expressions we use those of the first token of the left operand, the binary operator, and the first token of the right operand.
\cs{Does this not use the same heuristics as Deepbugs to decide which tokens to get embeddings for?}
Since the representations contain contextual information, the returned vectors can capture information about the rest of the tokens in the code sequence.

\section{Results}

We next discuss the experiments we performed and their corresponding results.
We measured the performance of the three baselines as well as those of non-contextual ELMO and SCELMO.
Measuring the performance of non-contextual ELMO allows us to evaluate how much improvement is due to specifics of the language model architecture, such as the character convolutional layer which can handle OOVs, and how much is due to the contextual information itself.
\cs{We need to explain what non-contextual ELMO and SCELMO is. How much of the source file
is SCELMO run on?}
\cs{We also need to explain what is the point of presenting results of no context ELMo. 
What scientific point does this make?}

\subsection{Performance on Validation Set}\label{sec:add-OOV}
In our first experiment we evaluate the performance of the methods in tasks where training data from the same projects are available.
The evaluation performed in this experiment gives a good estimation of how our method performs compared to the previous state-of-the-art technique of DeepBugs.
\cs{Please start by explaining what this experiment was designed to show.}
\cs{We should explain that this is run on the validation set, and what this means.}
One main difference however is that the evaluation now also includes instances which contain OOV.
As a consequence the bug detections tasks are harder than those presented by \cite{Pradel2018} as their evaluation does not include in both the training and validation set any instance for which an extracted identifier is OOV.
Table~\ref{tab:val-oov} illustrates the performance of the baselines and our models.
As one would expect the \ft\ baseline improves over \wVec\ for all bug types due to the subword information.
Moreover, our model SCELMo massively outperforms all other methods.
Lastly, even no-context ELMo the heuristic version of SCELMo that does not utilize contextual information at test time outperforms the baseline methods showcasing how powerful the pretrained representations are.

\begin{table*}[tb]
\centering
\caption{Comparison of ELMo versus non-contextual embeddings for bug detection on a validation set of projects.
Data is restricted to expressions that contain only single names.} \label{tab:val-oov}
\begin{tabular}[t]{l c c c c c}
\toprule
                      &    Random   &   \wVec   &  FastText  & No-Context ELMo & SCELMo \\
\midrule
Swapped Arguments     &    86.18\%  &  87.38\%  &   89.55\%  &    90.02\%        &   92.11\%    \\
Wrong Binary Operator &    90.47\%  &  91.05\%  &   91.11\%  &    92.47\%        &  100.00\%    \\
Wrong Binary Operand  &    75.56\%  &  77.06\%  &   79.74\%  &    81.71\%        &   84.23\%    \\
\bottomrule
\end{tabular}
\end{table*}

\subsection{Including Complex Expressions} \label{sec:complex-expr}
In our next experiment we also included instances that contain elements that are complex or nested expressions.
For instance, in the original work if one the arguments of a function call or one of the operands of a binary expression  is an expression consisting of other expressions then the instance would not be included in the dataset.
Several AST node types such as a \texttt{NewExpression} node or an \texttt{ObjectExpression} were not supported.
Figure~\ref{fig:complexExamples} a few examples of instances that would be previously skipped \footnote{The AST is extracted using the acorn parser \url{https://github.com/acornjs/acorn}}.
\cs{Please give either a citation or a URL for the Acorn parser.}
Such instances were skipped by \citet{Pradel2018} and not included in their results.
We do note though that we still skip very long expressions that contain more than 1000 tokens.

\begin{figure}[h]
\begin{center}

\begin{lstlisting}[]
// First argument is binary expression 
doComputation(x + find_min(components), callback);
\end{lstlisting}

\begin{lstlisting}[]
// Second argument is an unsupported node
factory.test(simulator, new Car('Eagle', 'Talon TSi', 1993));
\end{lstlisting}
\end{center}
\caption{Examples of instances that would be skipped by DeepBugs.}
\label{fig:complexExamples}
\end{figure}

Similarly to the previous experiment SCELMo significantly outperforms all other models.
This is evident in Table~\ref{tab:val-complex}.
Lastly, we clarify that the results of this section should not be directly compared to those of the previous one as for this experiment the training set is also larger.

\begin{table*}[tb] 
\centering
\caption{Comparison of SCELMo versus static embeddings on bug detection on a validation set of projects.
Complex expressions are included in this validation set.} \label{tab:val-complex}
\begin{tabular}[t]{l c c c c c}
\toprule
                      &    Random   &   \wVec   &  FastText  & No-Context ELMo &  SCELMo   \\
\midrule
Swapped Arguments     &    86.37\%  &  87.68\%  &   90.37\%  &    90.83\%        &   92.27\%   \\
Wrong Binary Operator &    91.12\%  &  91.68\%  &   91.92\%  &    92.75\%        &  100.00\%   \\
Wrong Binary Operand  &    72.73\%  &  74.31\%  &   77.41\%  &    79.65\%        &   87.10\%   \\
\bottomrule
\end{tabular}
\end{table*}

\subsection{External Test Evaluation}
The last experiment's objective is to showcase how the various models would perform on unseen projects as this better illustrates the generalizability of the techniques.
The configuration utilized is identical to that of the previous section.
By looking at Table~\ref{tab:test-res} one can notice that the baselines have a major drop in performance.
This is a common finding in machine learning models of code, namely, that applying a trained model to a new
software project is much more difficult than to a new file in the same project.
In contrast, SCELMo offers up to 15\% improvement in accuracy compared to \wVec\ baseline.
In fact, impressively enough SCELMo on the external test set is better than the evaluation set one of the baselines.

\begin{table*}[tb] 
\centering
\caption{Comparison of SCELMo versus static embeddings on bug detection on an external test set of 500 JavaScript projects.} \label{tab:test-res}
\begin{tabular}[t]{l c c c c c}
\toprule
                      &    Random   &   \wVec   &  FastText  & No-Context ELMo  & SCELMo  \\
\midrule
Swapped Arguments     &    75.79\%  &  78.22\%  &   79.40\%  &     81.37\%        &  84.25\% \\
Wrong Binary Operator &    82.95\%  &  85.54\%  &   83.15\%  &     86.54\%        &  99.99\% \\
Wrong Binary Operand  &    67.46\%  &  69.50\%  &   72.55\%  &     75.74\%        &  83.59\%  \\
\bottomrule
\end{tabular}
\end{table*}

\subsection{OOV Statistics}\label{sec:OOV-stats}

In order to better understand the above results we measured the OOV rate of the basic elements of the code instances appearing in the dataset.
Here the OOV rate is calculated based on the vocabulary of 10000 entries utilized by the \wVec and random baseline models.
These are illustrated in Tables~\ref{tab:call-stats}~and~\ref{tab:binOp-stats}.
We measured the OOV rates for both the version of the dataset used in Section~\ref{sec:OOV-stats},
which we call Train and Validation, and that used in Section~\ref{sec:complex-expr},
which we call Extended Train and Extended Validation.

Tables~\ref{tab:call-stats}~and~\ref{tab:binOp-stats} describe the OOV rates for different parts
of the expression types that are considered by the \DB\ bug detector.
A detailed description of the identifiers extraction heuristic can be found in Appendix~\ref{sec:extraction}.
\cs{This requires much more information. The reader will not be able to understand what any of the numbers
in this table mean. What does ``Calls missing base object'' mean? What does ``base object missing'' mean?}
\cs{Say in the text what the phrases on each row of the table means.}
We first focus on the swapped arguments bug pattern and consider all of the method call that have exactly two arguments.
Each method call contains the function name, a name of the first argument, a name of the second argument, and a base object.
The base object is the identifier that would be extracted from the expression (if such an expression exists) on which the function is called. 
For instance, from the following expression: \textit{window.navigator.userAgent.indexOf("Chrome")}, \textit{userAgent} would be extracted as the base object.
Table~\ref{tab:call-stats} shows for each of the components how often they are OOV.
In the expanded version of the dataset if one of the arguments is a complex expression then it is converted into a name based on the heuristic described in Section~\ref{sec:extraction}.
The resulting statistics contain valuable information as for instance, it is almost impossible for the \wVec\ baseline to reason about a swap arguments bug if the identifiers extracted for both arguments are OOV.

In a similar manner for the incorrect operand and operator bug patterns we consider all the binary operations.
Each binary expression consists of a left and right operand and a name is extracted for each of them.
For each operand we also measured the frequency with which the operand corresponds to certain common types such as identifier, literal or a \textit{ThisExpression}.

\begin{table*}[htb]
\centering
\caption{OOV statistics for calls with exactly two arguments (Swapped arguments instances). The statistics are calculated on variants of the DeepBugs dataset.}\label{tab:call-stats}
\begin{tabular}[t]{l c c c c}
\toprule
                            &      Train           &   Expanded Train   &         Validation     & Expanded Validation \\
\midrule
\textbf{Two Arguments Calls}&    \textbf{574656}   &   \textbf{888526}  &    \textbf{289061}     & \textbf{453486} \\
\midrule
Calls Missing Base Object   &       25.07\%        &       28.63\%      &         25.63\%        &      28.80\% \\
Base Object Missing or OOV  &       34.56\%        &       37.38\%      &         35.57\%        &      38.07\% \\
Function Name OOV           &       20.69\%        &       17.07\%      &         20.33\%        &      16.94\% \\
First Argument OOV          &       31.01\%        &       36.99\%      &         31.64\%        &      37.15\% \\
Second Argument OOV         &       27.25\%        &       22.86\%      &         27.94\%        &      23.49\% \\
Both Arguments OOV          &       11.33\%        &        9.57\%      &         11.96\%        &      10.16\% \\
Base and Function Name OOV  &       10.20\%        &        8.32\%      &         10.39\%        &       8.61\% \\
Base and Arguments OOV      &        4.21\%        &        3.31\%      &          4.88\%        &       3.77\% \\
Function Name and Arguments OOV    &        2.86\%        &        2.26\%      &          2.85\%        &       2.28\% \\
All Elements OOV            &        1.53\%        &        1.18\%      &          1.61\%        &       1.27\% \\
\bottomrule
\end{tabular}
\end{table*}

\begin{table*}[htb]
\centering
\caption{OOV statistics for binary operations.}\label{tab:binOp-stats}
\begin{tabular}[t]{l c c c c}
\toprule
                            &      Train           &   Expanded Train   &           Validation     & Expanded Validation \\
\midrule
\textbf{Binary Operations}  &   \textbf{1075175}   &  \textbf{1578776}  &     \textbf{540823}      & \textbf{797108} \\
\midrule
Left Operand OOV            &       25.40\%        &       28.84\%      &          26.04\%         &      29.55\% \\
Right Operand OOV           &       20.37\%        &       23.98\%      &          20.74\%         &      24.55\% \\
Both Operands OOV           &        7.82\%        &       11.29\%      &           8.24\%         &      11.88\% \\
Unknown Left Operand Type   &       83.36\%        &       87.80\%      &          83.14\%         &      87.74\% \\
Unknown Right Operand Type  &       48.48\%        &       47.23\%      &          48.47\%         &      47.05\% \\
Both Operand Types Unknown  &       33.34\%        &       36.06\%      &          33.20\%         &      35.87\% \\
All OOV or Unknown          &        3.59\%        &        4.03\%      &           3.81\%         &       4.3\% \\
\bottomrule
\end{tabular}
\end{table*}

\section{Is neural bug-finding useful in practice?}
Although related work \citep{Pradel2018, allamanis2018learning, Vasic2019neural} has shown that there is great potential for embedding based neural bug finders, the evaluation has mostly focused on synthetic bugs introduced by mutating the original code.
However, there is no strong indication that the synthetic bugs correlate to real ones, apart from a small study of the top 50 warnings for each bug type produced by DeepBugs.
A good example is the mutation operation utilized for the incorrect binary operator bug.
A lot of the introduced bug instances could result in syntactic errors.
This can potentially create a classifier with a high bias towards correlating buggy code to syntactically incorrect code, thus hindering the model's ability to generalize on real bugs.
Ideally, in an industrial environment we would like the resulting models to achieve a false positive rate of less than 10 \% \citep{Sadowski2015}.
Sadly, high true positive rates are not to be expected as well since static bug detectors were shown to be able to detect less than 5\% of bugs \citep{Habib2018} contained in the Defects4J corpus \citep{Just2014} and less than 12\% in a single-statement bugs corpus \citep{Karampatsis2019often}.
We note that in the second case the static analysis tool is given credit by reported any warning for the buggy line, so the actual percentage might lower than the reported one.

We next make a first step on investigating the practical usefulness of our methods by applying the classifiers of the previous section on a small corpus of real JavaScript bugs.
However, we think that this is a very hard yet interesting problem that should be carefully examined in future work.
In order to mine a corpus of real bug changes we used the methodology described in \citep{Karampatsis2019often}.
We note that we adapted their implementation to utilize the Rhino JavaScript parser\footnote{\url{https://github.com/mozilla/rhino}}.
Their methodology extracts bug fixing commits and filters them to only keep those that contain small single-statement changes.
Finally, it classifies each pair of modified statements by whether the fit a set of mutation patterns.
The resulting dataset is shown in Table~\ref{tab:real-bugs}.

\begin{table*}[tb] 
\centering
\caption{Real bug mined instances.} \label{tab:real-bugs}
\begin{tabular}[t]{l c c c c c}
\toprule
                      &    Swapped Arguments   &   Wrong Binary Operator   &  Wrong Binary Operand    \\
\midrule
Mined Instances       &           303          &              80           &           1007           \\
\bottomrule
\end{tabular}
\end{table*}

Finally, we queried the DeepBugs and SCELMo with each buggy instance as well as its fixed variant and measured the percentage of correctly classified instances for each of the two categories.
We also ignored any instances for which the JavaScript parser utilized for both failed to extract an AST.
We classified as bugs any instances that were assigned a probability to be a bug $> 75\%$.
In an actual system this threshold should ideally be tuned on a validation set.

\begin{table*}[tb] 
\centering
\caption{Real bug identification task recall and false positive rate (FPR).} \label{tab:bugs-res}
\begin{tabular}[t]{l c c c c}
\toprule
                      &    \wVec-Recall  &   \wVec-FPR   &  SCELMo-Recall  &  SCELMo-FPR  \\
\midrule
Swapped Arguments     &     3.34\%     &      0.33\%    &    49.67\%    &     33.78\%    \\
Wrong Binary Operator &     8.95\%     &       7.70\%    &     0.00\%    &    0.00\%    \\
Wrong Binary Operand  &    11.99\%     &      12.11\%    &    15.81\%    &    14.34\%    \\
\bottomrule
\end{tabular}
\end{table*}

Table~\ref{tab:bugs-res} suggests that there might indeed be some potential for future practical applications of neural bug finding techniques.
Both are able to uncover some of the bugs.
However, the results also suggest that careful tuning of the predictions threshold might be necessary, especially if we take into account the industrial need to comply with a low false positive rate (FPR).
For instance, raising SCELMo's prediction threshold to $80\%$ for the swap arguments bug results in finding only 3.34\% of the bugs but correctly classifying 100\% of the repaired function calls, thus achieving 0.0\% false positive rate.
Moreover, since SCELMo could not uncover any of the real binary operator bugs, future work could investigate the effect of utilizing different mutation strategies for the purpose of artificial bug-induction.
Future work could also investigate if fine-tuning on small set of real bugs could result in more robust classifiers.

\section{Conclusion}

We have presented SCELMo, which is to our knowledge the first language-model based
contextual embeddings for source code. Contextual embeddings have many potential advantages
for source code, because surrounding tokens can indirectly provide information
about tokens, e.g. about likely values of variables. We highlight the utility
of SCELMo embeddings by using them within a recent state-of-the-art machine
learning based bug detector. The SCELMo embeddings yield a dramatic improvement
in the synthetic bug detection performance benchmark, especially on lines of code that contain out-of-vocabulary
tokens and complex expressions that can cause difficulty for the method.
We also showed and discussed the performance of the resulting bug detectors on a dataset of real bugs raising useful insights for future work.

\subsubsection*{Acknowledgements}
This work was supported in part by the EPSRC Centre for Doc-toral Training in Data Science, funded by the UK EngineeringandPhysical Sciences Research Council (grant EP/L016427/1) and the University of Edinburgh.

\pagebreak

\bibliography{iclr2020_conference}

\begin{thebibliography}{31}
\providecommand{\natexlab}[1]{#1}
\providecommand{\url}[1]{\texttt{#1}}
\expandafter\ifx\csname urlstyle\endcsname\relax
  \providecommand{\doi}[1]{doi: #1}\else
  \providecommand{\doi}{doi: \begingroup \urlstyle{rm}\Url}\fi

\bibitem[Allamanis et~al.(2018{\natexlab{a}})Allamanis, Barr, Devanbu, and
  Sutton]{Allamanis2018}
Miltiadis Allamanis, Earl~T. Barr, Premkumar Devanbu, and Charles Sutton.
\newblock A survey of machine learning for big code and naturalness.
\newblock \emph{ACM Comput. Surv.}, 51\penalty0 (4):\penalty0 81:1--81:37, July
  2018{\natexlab{a}}.
\newblock ISSN 0360-0300.
\newblock \doi{10.1145/3212695}.
\newblock URL \url{http://doi.acm.org/10.1145/3212695}.

\bibitem[Allamanis et~al.(2018{\natexlab{b}})Allamanis, Brockschmidt, and
  Khademi]{allamanis2018learning}
Miltiadis Allamanis, Marc Brockschmidt, and Mahmoud Khademi.
\newblock Learning to represent programs with graphs.
\newblock In \emph{International Conference on Learning Representations},
  2018{\natexlab{b}}.
\newblock URL \url{https://openreview.net/forum?id=BJOFETxR-}.

\bibitem[Alon et~al.(2019)Alon, Zilberstein, Levy, and Yahav]{Alon2019}
Uri Alon, Meital Zilberstein, Omer Levy, and Eran Yahav.
\newblock Code2vec: Learning distributed representations of code.
\newblock \emph{Proc. ACM Program. Lang.}, 3\penalty0 (POPL):\penalty0
  40:1--40:29, January 2019.
\newblock ISSN 2475-1421.
\newblock \doi{10.1145/3290353}.
\newblock URL \url{http://doi.acm.org/10.1145/3290353}.

\bibitem[Bavishi et~al.(2018)Bavishi, Pradel, and Sen]{Bavishi2018}
Rohan Bavishi, Michael Pradel, and Koushik Sen.
\newblock Context2name: {A} deep learning-based approach to infer natural
  variable names from usage contexts.
\newblock \emph{CoRR}, abs/1809.05193, 2018.
\newblock URL \url{http://arxiv.org/abs/1809.05193}.

\bibitem[Bojanowski et~al.(2017)Bojanowski, Grave, Joulin, and
  Mikolov]{Bojanowski2017}
Piotr Bojanowski, Edouard Grave, Armand Joulin, and Tomas Mikolov.
\newblock Enriching word vectors with subword information.
\newblock \emph{Transactions of the Association for Computational Linguistics},
  5:\penalty0 135--146, 2017.
\newblock \doi{10.1162/tacl_a_00051}.
\newblock URL \url{https://www.aclweb.org/anthology/Q17-1010}.

\bibitem[Chen \& Monperrus(2019)Chen and Monperrus]{Chen2019}
Zimin Chen and Martin Monperrus.
\newblock A literature study of embeddings on source code.
\newblock \emph{CoRR}, abs/1904.03061, 2019.
\newblock URL \url{http://arxiv.org/abs/1904.03061}.

\bibitem[Chen et~al.(2019)Chen, Kommrusch, Tufano, Pouchet, Poshyvanyk, and
  Monperrus]{Chen2019Sequencer}
Zimin Chen, Steve Kommrusch, Michele Tufano, Louis{-}No{\"{e}}l Pouchet, Denys
  Poshyvanyk, and Martin Monperrus.
\newblock Sequencer: Sequence-to-sequence learning for end-to-end program
  repair.
\newblock \emph{CoRR}, abs/1901.01808, 2019.
\newblock URL \url{http://arxiv.org/abs/1901.01808}.

\bibitem[Devlin et~al.(2018)Devlin, Chang, Lee, and Toutanova]{Devlin2018}
Jacob Devlin, Ming{-}Wei Chang, Kenton Lee, and Kristina Toutanova.
\newblock {BERT:} pre-training of deep bidirectional transformers for language
  understanding.
\newblock \emph{CoRR}, abs/1810.04805, 2018.
\newblock URL \url{http://arxiv.org/abs/1810.04805}.

\bibitem[Gazzola et~al.(2019)Gazzola, Micucci, and Mariani]{Gazzola2019}
L.~Gazzola, D.~Micucci, and L.~Mariani.
\newblock Automatic software repair: A survey.
\newblock \emph{IEEE Transactions on Software Engineering}, 45\penalty0
  (01):\penalty0 34--67, jan 2019.
\newblock ISSN 1939-3520.
\newblock \doi{10.1109/TSE.2017.2755013}.

\bibitem[Habib \& Pradel(2018)Habib and Pradel]{Habib2018}
Andrew Habib and Michael Pradel.
\newblock How many of all bugs do we find? a study of static bug detectors.
\newblock In \emph{Proceedings of the 33rd ACM/IEEE International Conference on
  Automated Software Engineering}, ASE 2018, pp.\  317--328, New York, NY, USA,
  2018. ACM.
\newblock ISBN 978-1-4503-5937-5.
\newblock \doi{10.1145/3238147.3238213}.
\newblock URL \url{http://doi.acm.org/10.1145/3238147.3238213}.

\bibitem[Harer et~al.(2018)Harer, Kim, Russell, Ozdemir, Kosta, Rangamani,
  Hamilton, Centeno, Key, Ellingwood, McConley, Opper, Chin, and
  Lazovich]{Harer2018}
Jacob~A. Harer, Louis~Y. Kim, Rebecca~L. Russell, Onur Ozdemir, Leonard~R.
  Kosta, Akshay Rangamani, Lei~H. Hamilton, Gabriel~I. Centeno, Jonathan~R.
  Key, Paul~M. Ellingwood, Marc~W. McConley, Jeffrey~M. Opper, Sang~Peter Chin,
  and Tomo Lazovich.
\newblock Automated software vulnerability detection with machine learning.
\newblock \emph{CoRR}, abs/1803.04497, 2018.
\newblock URL \url{http://arxiv.org/abs/1803.04497}.

\bibitem[Just et~al.(2014)Just, Jalali, and Ernst]{Just2014}
Ren{\'e} Just, Darioush Jalali, and Michael~D. Ernst.
\newblock Defects4j: A database of existing faults to enable controlled testing
  studies for java programs.
\newblock In \emph{Proceedings of the 2014 International Symposium on Software
  Testing and Analysis}, ISSTA 2014, pp.\  437--440, New York, NY, USA, 2014.
  ACM.
\newblock ISBN 978-1-4503-2645-2.
\newblock \doi{10.1145/2610384.2628055}.
\newblock URL \url{http://doi.acm.org/10.1145/2610384.2628055}.

\bibitem[Karampatsis \& Sutton(2019)Karampatsis and
  Sutton]{Karampatsis2019often}
Rafael-Michael Karampatsis and Charles Sutton.
\newblock {How Often Do Single-Statement Bugs Occur? The ManySStuBs4J Dataset}.
\newblock \emph{arXiv preprint arXiv:1905.13334}, 2019.
\newblock URL \url{https://arxiv.org/abs/1905.13334}.

\bibitem[Li et~al.(2016)Li, Zemel, Brockschmidt, and Tarlow]{Li2016gated}
Yujia Li, Richard Zemel, Marc Brockschmidt, and Daniel Tarlow.
\newblock Gated graph sequence neural networks.
\newblock In \emph{Proceedings of ICLR'16}, April 2016.
\newblock URL
  \url{https://www.microsoft.com/en-us/research/publication/gated-graph-sequence-neural-networks/}.

\bibitem[Liu et~al.(2019)Liu, Ott, Goyal, Du, Joshi, Chen, Levy, Lewis,
  Zettlemoyer, and Stoyanov]{Liu2019}
Yinhan Liu, Myle Ott, Naman Goyal, Jingfei Du, Mandar Joshi, Danqi Chen, Omer
  Levy, Mike Lewis, Luke Zettlemoyer, and Veselin Stoyanov.
\newblock Roberta: {A} robustly optimized {BERT} pretraining approach.
\newblock \emph{CoRR}, abs/1907.11692, 2019.
\newblock URL \url{http://arxiv.org/abs/1907.11692}.

\bibitem[Long \& Rinard(2016)Long and Rinard]{Long2016}
Fan Long and Martin Rinard.
\newblock Automatic patch generation by learning correct code.
\newblock In \emph{Proceedings of the 43rd Annual ACM SIGPLAN-SIGACT Symposium
  on Principles of Programming Languages}, POPL '16, pp.\  298--312, New York,
  NY, USA, 2016. ACM.
\newblock ISBN 978-1-4503-3549-2.
\newblock \doi{10.1145/2837614.2837617}.
\newblock URL \url{http://doi.acm.org/10.1145/2837614.2837617}.

\bibitem[Melamud et~al.(2016)Melamud, Goldberger, and Dagan]{Melamud2016}
Oren Melamud, Jacob Goldberger, and Ido Dagan.
\newblock context2vec: Learning generic context embedding with bidirectional
  {LSTM}.
\newblock In \emph{Proceedings of The 20th {SIGNLL} Conference on Computational
  Natural Language Learning}, pp.\  51--61, Berlin, Germany, August 2016.
  Association for Computational Linguistics.
\newblock \doi{10.18653/v1/K16-1006}.
\newblock URL \url{https://www.aclweb.org/anthology/K16-1006}.

\bibitem[Mikolov et~al.(2013)Mikolov, Sutskever, Chen, Corrado, and
  Dean]{Mikolov2013}
Tomas Mikolov, Ilya Sutskever, Kai Chen, Greg~S Corrado, and Jeff Dean.
\newblock Distributed representations of words and phrases and their
  compositionality.
\newblock In C.~J.~C. Burges, L.~Bottou, M.~Welling, Z.~Ghahramani, and K.~Q.
  Weinberger (eds.), \emph{Advances in Neural Information Processing Systems
  26}, pp.\  3111--3119. Curran Associates, Inc., 2013.
\newblock URL
  \url{http://papers.nips.cc/paper/5021-distributed-representations-of-words-and-phrases-and-their-compositionality.pdf}.

\bibitem[Monperrus(2018)]{Monperrus2018}
Martin Monperrus.
\newblock Automatic software repair: A bibliography.
\newblock \emph{ACM Comput. Surv.}, 51\penalty0 (1):\penalty0 17:1--17:24,
  January 2018.
\newblock ISSN 0360-0300.
\newblock \doi{10.1145/3105906}.
\newblock URL \url{http://doi.acm.org/10.1145/3105906}.

\bibitem[Neelakantan et~al.(2014)Neelakantan, Shankar, Passos, and
  McCallum]{Neelakantan2014}
Arvind Neelakantan, Jeevan Shankar, Alexandre Passos, and Andrew McCallum.
\newblock Efficient non-parametric estimation of multiple embeddings per word
  in vector space.
\newblock In \emph{Proceedings of the 2014 Conference on Empirical Methods in
  Natural Language Processing ({EMNLP})}, pp.\  1059--1069, Doha, Qatar,
  October 2014. Association for Computational Linguistics.
\newblock \doi{10.3115/v1/D14-1113}.
\newblock URL \url{https://www.aclweb.org/anthology/D14-1113}.

\bibitem[Pennington et~al.(2014)Pennington, Socher, and
  Manning]{Pennington2014}
Jeffrey Pennington, Richard Socher, and Christopher Manning.
\newblock {G}love: Global vectors for word representation.
\newblock In \emph{Proceedings of the 2014 Conference on Empirical Methods in
  Natural Language Processing ({EMNLP})}, pp.\  1532--1543, Doha, Qatar,
  October 2014. Association for Computational Linguistics.
\newblock \doi{10.3115/v1/D14-1162}.
\newblock URL \url{https://www.aclweb.org/anthology/D14-1162}.

\bibitem[Peters et~al.(2018)Peters, Neumann, Iyyer, Gardner, Clark, Lee, and
  Zettlemoyer]{Peters2018}
Matthew Peters, Mark Neumann, Mohit Iyyer, Matt Gardner, Christopher Clark,
  Kenton Lee, and Luke Zettlemoyer.
\newblock Deep contextualized word representations.
\newblock In \emph{Proceedings of the 2018 Conference of the North {A}merican
  Chapter of the Association for Computational Linguistics: Human Language
  Technologies, Volume 1 (Long Papers)}, pp.\  2227--2237, New Orleans,
  Louisiana, June 2018. Association for Computational Linguistics.
\newblock \doi{10.18653/v1/N18-1202}.
\newblock URL \url{https://www.aclweb.org/anthology/N18-1202}.

\bibitem[Pradel \& Sen(2018)Pradel and Sen]{Pradel2018}
Michael Pradel and Koushik Sen.
\newblock Deepbugs: A learning approach to name-based bug detection.
\newblock \emph{Proc. ACM Program. Lang.}, 2\penalty0 (OOPSLA):\penalty0
  147:1--147:25, October 2018.
\newblock ISSN 2475-1421.
\newblock \doi{10.1145/3276517}.
\newblock URL \url{http://doi.acm.org/10.1145/3276517}.

\bibitem[Raychev et~al.(2016)Raychev, Bielik, Vechev, and Krause]{Raychev2016}
Veselin Raychev, Pavol Bielik, Martin Vechev, and Andreas Krause.
\newblock Learning programs from noisy data.
\newblock In \emph{Proceedings of the 43rd Annual ACM SIGPLAN-SIGACT Symposium
  on Principles of Programming Languages}, POPL '16, pp.\  761--774, New York,
  NY, USA, 2016. ACM.
\newblock ISBN 978-1-4503-3549-2.
\newblock \doi{10.1145/2837614.2837671}.
\newblock URL \url{http://doi.acm.org/10.1145/2837614.2837671}.

\bibitem[Sadowski et~al.(2015)Sadowski, van Gogh, Jaspan, S\"{o}derberg, and
  Winter]{Sadowski2015}
Caitlin Sadowski, Jeffrey van Gogh, Ciera Jaspan, Emma S\"{o}derberg, and
  Collin Winter.
\newblock Tricorder: Building a program analysis ecosystem.
\newblock In \emph{Proceedings of the 37th International Conference on Software
  Engineering - Volume 1}, ICSE '15, pp.\  598--608, Piscataway, NJ, USA, 2015.
  IEEE Press.
\newblock ISBN 978-1-4799-1934-5.
\newblock URL \url{http://dl.acm.org/citation.cfm?id=2818754.2818828}.

\bibitem[Turian et~al.(2010)Turian, Ratinov, and Bengio]{Turian2010}
Joseph Turian, Lev Ratinov, and Yoshua Bengio.
\newblock Word representations: A simple and general method for semi-supervised
  learning.
\newblock In \emph{Proceedings of the 48th Annual Meeting of the Association
  for Computational Linguistics}, ACL '10, pp.\  384--394, Stroudsburg, PA,
  USA, 2010. Association for Computational Linguistics.
\newblock URL \url{http://dl.acm.org/citation.cfm?id=1858681.1858721}.

\bibitem[Vasic et~al.(2019)Vasic, Kanade, Maniatis, Bieber, and
  singh]{Vasic2019neural}
Marko Vasic, Aditya Kanade, Petros Maniatis, David Bieber, and Rishabh singh.
\newblock Neural program repair by jointly learning to localize and repair.
\newblock In \emph{International Conference on Learning Representations}, 2019.
\newblock URL \url{https://openreview.net/forum?id=ByloJ20qtm}.

\bibitem[White et~al.(2019)White, Tufano, Martinez, Monperrus, and
  Poshyvanyk]{White2019}
Martin White, Michele Tufano, Matias Martinez, Martin Monperrus, and Denys
  Poshyvanyk.
\newblock Sorting and transforming program repair ingredients via deep learning
  code similarities.
\newblock pp.\  479--490, 02 2019.
\newblock \doi{10.1109/SANER.2019.8668043}.

\bibitem[Wieting et~al.(2016)Wieting, Bansal, Gimpel, and Livescu]{Wieting2016}
John Wieting, Mohit Bansal, Kevin Gimpel, and Karen Livescu.
\newblock {C}haragram: Embedding words and sentences via character n-grams.
\newblock In \emph{Proceedings of the 2016 Conference on Empirical Methods in
  Natural Language Processing}, pp.\  1504--1515, Austin, Texas, November 2016.
  Association for Computational Linguistics.
\newblock \doi{10.18653/v1/D16-1157}.
\newblock URL \url{https://www.aclweb.org/anthology/D16-1157}.

\bibitem[Yang et~al.(2019)Yang, Dai, Yang, Carbonell, Salakhutdinov, and
  Le]{Yang2019}
Zhilin Yang, Zihang Dai, Yiming Yang, Jaime~G. Carbonell, Ruslan Salakhutdinov,
  and Quoc~V. Le.
\newblock {XLNet}: Generalized autoregressive pretraining for language
  understanding.
\newblock \emph{CoRR}, abs/1906.08237, 2019.
\newblock URL \url{http://arxiv.org/abs/1906.08237}.

\bibitem[Yin et~al.(2018)Yin, Neubig, Allamanis, Brockschmidt, and
  Gaunt]{Yin2018}
Pengcheng Yin, Graham Neubig, Miltiadis Allamanis, Marc Brockschmidt, and
  Alexander~L. Gaunt.
\newblock Learning to represent edits.
\newblock \emph{CoRR}, abs/1810.13337, 2018.
\newblock URL \url{http://arxiv.org/abs/1810.13337}.

\end{thebibliography}
\bibliographystyle{iclr2020_conference}
\pagebreak

\appendix
\section{Name Extraction Heuristic} \label{sec:extraction}
In order for DeepBugs to operate it is necessary to extract identifiers or literals for each expression part of the statement.
The bug detector for swapped arguments utilizes the following elements of the function call:
\begin{description}
\item[Base Object:] The expression on which the function is called.
\item[Callee:] The called function.
\item[Argument 1:] The expression consisting the first argument of the called function.
\item[Argument 2:] The expression consisting the first argument of the called function.
\end{description}

Similarly the bug detectors for incorrect binary operators and operands utilize the following elements of the binary expression:
\begin{description}
\item[Binary Operator:] The binary operator utilized in the expression.
\item[Left Operand:] The left operand of the binary expression.
\item[Right Operand:] The right operand of the binary expression.
\end{description}

We next describe the extraction heuristic, which is shared by all the bug detectors.
The heuristic takes as input a node $n$ representing an expression and returns $name(n)$ based on the following rules:
\begin{itemize}
\item Identifier: return its name.
\item Literal: return its value.
\item this expression: return \textit{this}.
\item Update expression with argument $x$: return $name(x)$.
\item Member expression accessing a property $p$: return $name(p)$.
\item Member expression accessing a property $base[p]$: return $name(base)$.
\item Call expression $base.callee(...)$: return $name(callee)$.
\item Property node $n$: If $n.key$ does not exist return $name(n.value)$. If $name(n.key)$ does not exist return $name(n.value)$ . Otherwise randomly return either $name(n.value)$ or $name()n.key)$.
\item Binary expression with left operand $l$ and right operand $r$: Run the heuristic on both $l$ and $r$ to retrieve $name(l)$ and $name(r)$. If $name(l)$ does not exist return $name(r)$. If $name(r)$ does not exist return $name(l)$. Otherwise randomly return either $name(l)$ ir $name(r)$.
\item Logical expression with left operand $l$ and right operand $r$: Run the heuristic on both $l$ and $r$ to retrieve $name(l)$ and $name(r)$. If $name(l)$ does not exist return $name(r)$. If $name(r)$ does not exist return $name(l)$. Otherwise randomly return either $name(l)$ ir $name(r)$.
\item Assignment expression with left operand $l$ and right operand $r$: Run the heuristic on both $l$ and $r$ to retrieve $name(l)$ and $name(r)$. If $name(l)$ does not exist return $name(r)$. If $name(r)$ does not exist return $name(l)$. Otherwise, randomly return either $name(l)$ ir $name(r)$.
\item Unary expression with argument $u$ : Return $name(u)$.
\item Array expression with elements $l_i$ : For all $l_i$ that $name(l_i)$ exists randomly choose one of them  and return $name(l_i)$.
\item Conditional expression with operands $c$, $l$, and $r$: Randomly choose one out of $c$, $l$, $r$ for which a name exists and return its name. 
\item Function expression: return $function$.
\item Object expression: return $\lbrace$.
\item New expression with a constructor function call $c$: return $name(c)$.
\end{itemize}
All random decisions follow a uniform distribution.

\end{document}